\documentclass[twocolumn,superscriptaddress,preprintnumbers,amsmath,amssymb]{revtex4}
\usepackage{graphicx}
\usepackage{dcolumn}
\usepackage{bm}
\newcommand{\lh}[1]{\slash \! \! \!  #1}

\begin{document}

\title{A new determination of the electromagnetic nucleon form factors from QCD Sum Rules}
\author{H. Castillo}
\affiliation{Departamento de Ciencias, Pontificia Universidad
Cat\'{o}lica del
 Per\'u, Apartado 1761, Lima, Per\'u.}
 \author{C. A. Dominguez}
\affiliation{Institute of Theoretical Physics \& Astrophysics,
University of Cape Town, Rondebosch 7701, South Africa.}
\author{M. Loewe}
\affiliation{Facultad de F\'\i sica, Pontificia Universidad
Cat\'olica de Chile,\\ Casilla 306, Santiago 22, Chile.}

\begin{abstract}
We obtain the electromagnetic form factors of the nucleon, in the
space-like region, using three-point function Finite Energy QCD
Sum Rules. The QCD calculation is performed to leading order in
perturbation theory in the chiral limit, and also to leading order
in the non-perturbative power corrections. For the Dirac form
factor, $F_1(q^2)$, we get a  very good agreement with the data
for both the proton and the neutron, in the currently accessible
experimental region of momentum transfers. Unfortunately this is
not the case, though, for the Pauli form factor $F_2(q^2)$, which
has a soft $q^2$-dependence proportional to the quark condensate
$<0|\bar{q}q|0>$.
\end{abstract}

\maketitle

The determination of the electromagnetic nucleon form factors is
an old standing problem in QCD. For a review, see \cite{REV1}.
Calculations based on perturbative QCD (PQCD), together with sum
rules estimates for the nucleon wave function, are difficult to
compare with data due to the extreme asymptotic nature of these
theoretical results. Recently, a new analysis based on light-cone
QCD Sum Rules \cite{BRAUN} has been carried out improving the
agreement with data from within a factor 5-6 to a factor of two.
Here we attempt a Finite Energy QCD Sum Rules (FESR) determination
of the Dirac $F_{1}(Q^{2})$ and of the Pauli $F_{2}(Q^{2})$ form
factors, in the region of experimentally accessible momentum
transfers. The QCD-FESR approach is interesting, since power
corrections associated to vacuum condensates of different
dimensions decouple at leading order in PQCD.

 As it is well known this technique is based on the
Operator Product Expansion (OPE) of current correlators and on the
notion of quark-hadron duality \cite{QCDSR}. Our calculation will
be done to leading order in PQCD, in the chiral limit including
also the leading-order non-perturbative power corrections
associated to the quark-condensate and to the four-quark
condensate.

By considering the interpolating current with proton quantum
numbers
\begin{equation}
\eta_N(x)=\varepsilon_{abc}\left[u^a(x)(C\gamma_\alpha)u^b(x)\right]
(\gamma^5\gamma^\alpha d^c(x)), \label{ec1}
\end{equation}
\noindent and the electromagnetic current
\begin{equation}
J_{EM}^\mu (y)=\frac 2 3 \bar u(y)\gamma^\mu u(y)-\frac 1 3 \bar
d(y)\gamma^\mu d(y) \; ,
\end{equation}
\noindent we are interested in the three-point correlator
\begin{eqnarray}
\Pi_\mu (p^2,p'^{2},Q^2) =  i^2 \!\int \!\!\mbox{d}^ 4 x \!\!\int
 \!\!\mbox{d}^4 y \;
 e
^{i (p' \cdot x - q  \cdot y)} \nonumber\\
  \langle 0 \left| T \{ \eta_N
(x)
 J_\mu^{EM}(y)\bar
 \eta_N
(0) \} \right| 0\rangle \; ,
\end{eqnarray}
\noindent where $Q^{2} \equiv -q^{2} = -(p' - p)^{2} \geq 0$ is
fixed. See Fig.1. The current Eq.1 couples to a nucleon of
momentum $p$ and polarization $s $ according to
\begin{equation}
\langle 0\left| \eta_{N}(0) \right|N(p,s)\rangle = \lambda_N
u(p,s) ,
\end{equation}
\noindent where u(p,s) is a nucleon spinor and $\lambda _{N}$ is a
phenomenological parameter that gives us the current-nucleon
coupling. This parameter has been estimated, for example, using
two-point QCD sum rules involving the current $\eta _{N}$
\cite{reinders}-\cite{MNDL}.

Going to the hadronic sector, after inserting a one-particle
nucleon state, the three-point function (3) can be written in
terms of the nucleon form factors $F_{1}(Q^{2})$ and
$F_{2}(Q^{2})$, defined as
\begin{eqnarray}
\langle k_1\, s_1\left|  J_\mu^{EM}(0) \right| k_2\, s_2\rangle \
=
 \bar u_N (k_1,s_1) \nonumber\\
 \times \left[ F_1 (q^2) \gamma _\mu +
\frac{i \kappa}{2M_N} F_2(q^2)\sigma_{\mu \nu} q^\nu \right]
u_N(k_2,s_2) \; ,
\end{eqnarray}
\noindent where $q^2= (k_2-k_1)^2$, and $\kappa$ is the anomalous
magnetic moment in units of nuclear magnetons ($\kappa_p = 1.79$
for  the proton, and $\kappa_n = - 1.91$  for the neutron). The
form factors $F_{1,2}(q^2)$ are related to the electric and
magnetic (Sachs) form factors $G_E(q^2)$, and $G_M(q^2)$, measured
in elastic electron-proton scattering experiments, according to
\begin{eqnarray}
G_E(q^2) &\equiv& F_1(q^2)+\frac{\kappa q^2}{(2m)^2} F_2(q^2)\; ,
\\ [.2cm]
G_M(q^2) &\equiv& F_1(q^2)+ \kappa F_2(q^2) \; ,
\end{eqnarray}
where $G_E^p(0) = 1$, $G_M^p(0) = 1 + \kappa_p$ for the proton,
and $G_E^n(0) = 0$,  $G_M^n(0) = \kappa_n$ for the neutron.
 Next we compute the hadronic spectral function by
inserting a complete set of intermediate nucleonic states in (3)
and computing the double discontinuity in the complex $p^{2}
\equiv s$, $p'^{2} \equiv s'$ plane. If we stay with $s,s' < 2.1
GeV^{2}$, i.e. below the Roper resonance, we can approximate the
hadronic spectral function by the single-particle nucleon pole
plus a continuum with thresholds $s_{0}$ and $s'_{0}$ ($s_{0},
s'_{0} > M_{N}^{2}$) that we expect will coincide with the PQCD
spectral function (local duality). In this way we get
\begin{eqnarray}
\mbox{Im} \Pi_\mu(s,s',Q^2)|_{HAD}=\pi^2\,\lambda_N^2\, \delta
(s-M_N^2) \nonumber\\
\times \delta(s' - M_N^2)
 \{ F_1(q^2)
A_{\mu } + \frac{i \kappa}{2 M_N}F_2(q^2)B_{\mu \nu}\nonumber\\
q^\nu
\} \Theta\left(s_0-s\right)\nonumber\\
+ \; \mbox{Im} \Pi_{\mu}(s,s',Q^2)\Big|_{PQCD}\Theta\left(
s-s_0\right) \; ,
\end{eqnarray}
where for simplicity we set $s_0=s'_0$ and $A_{\mu}$ and $B_{\mu
\nu }$ correspond to the following tensor structures
\begin{eqnarray}
A_{\mu } = \lh p' \gamma_\mu \lh p+M_N (\lh p' \gamma_\mu+
\gamma_\mu \lh p)+  M_N^2 \gamma_\mu
\end{eqnarray}
\noindent and
\begin{eqnarray}
B_{\mu \nu} = \lh p' \sigma_{\mu\nu}\lh p+M_N(\lh p'
\sigma_{\mu\nu}+\sigma_{\mu\nu}\lh p)+M_N^2 \sigma_{\mu\nu}.
\end{eqnarray}
Going to the QCD sector, to leading order in PQCD and in the
chiral limit, we have to calculate the imaginary part of the
diagram shown in Fig.1.
\begin{figure}
\includegraphics[scale=0.5]{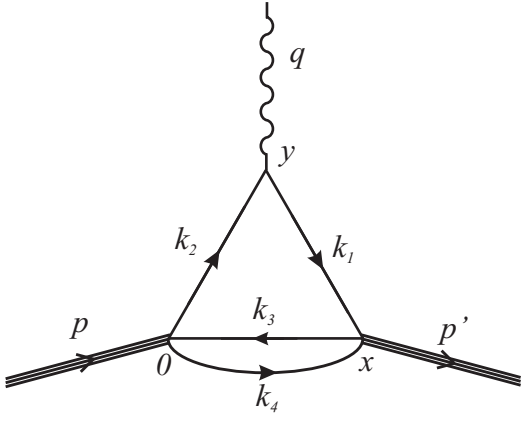}
\caption{The three-point function, eq. 3, to leading order in
PQCD} \label{fig1}
\end{figure}
The important point is that there are several Lorentz structures,
analogous to those we found in the hadronic sector. Before
invoking local duality it is necessary to choose a particular
Lorentz structure present both in the QCD as well as in the
hadronic sectors.

The term $ \lh p' \gamma_\mu \lh p$ turns out to be appropriate.
It allows to project $F_{1}(Q^{2})$ since this structure does not
appear together with $F_{2}(Q^{2})$ in the hadronic spectral
function. On the other hand, due to vanishing traces, the quark
condensate to be considered later, also does not involve this
structure. In principle, however, there are four quark condensate
terms associated with such structure. However, those terms do not
contribute to the FESR since the associate double discontinuity
vanishes. After a very lengthy calculation, the imaginary part of
the perturbative expression of the correlator, associated to the
desired structure $\lh{p'} \gamma_\mu \lh{p}$, can be written as

\begin{eqnarray}
\mbox{Im} \Pi^\mu(s,s',Q^2)= \lh p'\gamma^\mu \lh p
\;\left( \alpha + \beta [ -323\,Q^{12} - \right.\nonumber\\
Q^{10}P_{10}(s,s') - 10\,Q^8 P_{8}(s,s') +  Q^6 P_{6}(s,s')
 \nonumber\\
+ \; Q^4 P_{4}(s,s')  -  Q^2 P_{2}(s,s') - P_{0}(s,s')]) +\ldots
\end{eqnarray}
\noindent In the equation above the dots denote the terms
associated to other Lorentz structures and we have introduced
\begin{eqnarray}
\beta ^{-1} \! \! \! \! & = & \! \! \! 4608 \pi ^{2}[q^{4} + \!
(s-s')^{2} +
2Q^{2}(s+s')]^{\frac{5}{2}} \nonumber\\
\alpha \! \! \! & = & \! \! \! \frac{323\, Q^{2} +378\,
(s-s')}{4608 \pi^{2},}
\end{eqnarray}
\noindent and the  set of polynomials $P_{i}(s,s')$ given by

\smallskip
\noindent $P_{10} = 1993s +1237s'$,

\smallskip
\noindent $P_{8}= 512s^{2} + 323ss' + 134s'^{2}$,

\smallskip
\noindent $P_{6} = -7010 s^{3} +1188 s s'^{2} +550 s'^{3}$,

\smallskip
\noindent $P_{4} = -5395s^{4} + 7010 s^{3}s' + 2610 s^{2}s'^{2} +
3146s s'^{3} +  2165s'^{4}$

\smallskip
\noindent $P_{2} = (s-s')^{2}(2213 s^{3} - 2589 s^{2}s' - 3099 s
s'^{2} - 1567 s'^{3})$

\smallskip
\noindent $P_{0} = -378(s-s')^{4}(s^{2} - 2ss' -s'^{2})$

It is interesting to mention that we got both explicit terms,
where the desired tensor structure was there from the beginning,
as well as implicit terms, i.e. those terms where the tensor
structure emerged only after performing the integrals. The next
step is to invoke global quark-hadron duality in the frame of the
FESR, which, as we mentioned, are organized according to
dimensionality. The FESR of leading dimensionality are
\begin{eqnarray}
\int_0^{s_0}\mbox{d} s \int_0^{s_0-s} \mbox{d}
{s'}\;\mbox{Im}\Pi(s,s',Q^2)\mid_{HAD}= \nonumber\\
 \int_0^{s_0}
\mbox{d} s \int_0^{s_0-s} \mbox{d} {s'}\;
\mbox{Im}\Pi(s,s',Q^2)\mid_{QCD} \; .
\end{eqnarray}

We have chosen a triangular region to integrate in the $s,s'$
plane, but the result is quite independent from the integration
region \cite{IOFFE81} and \cite{DLR}. In this way one obtains
\begin{eqnarray}
F_1(Q^2)=\frac{1} {9216\,{\pi }^4\,( Q^2 + 2\,{s_0} ) \,
    {{{\lambda }_N}}^2} \nonumber \\
 \times (A + B\ln (\frac{Q^{2}}{Q^{2}+2s_{0}})),
\end{eqnarray}

\noindent where we have defined

\smallskip
\noindent $A = 2\,{s_0}( 96\,Q^6 + 297\,Q^4\,{s_0} + 158\,Q^2\,
{{s_0}}^2 - 112\,{{s_0}}^3 )$ and

\smallskip
\noindent $B = 3(Q^2  \! + \! 2\,{s_0} ) \, ( 32\,Q^6 \! +
67Q^4{s_0} + 7Q^2{{s_0}}^2 - 21{{s_0}}^3
 ). $

 \smallskip
 Notice that in the previous equation we have the
 standard logarithmic singularity arising from the chiral limit.
 The leading asymptotic term turns out to be
 \begin{equation}
   \lim_{Q^2 \rightarrow \infty}
        Q^4 \; F_1(Q^2) = \frac{11 \, s_0^5}{2560 \, \pi^4  \,
        \lambda_N^2} \; .
\end{equation}

Qualitatively, this asymptotic behaviour agrees with expectations.
From QCD sum rules for two-point functions involving the nucleon
current (1) it has been found \cite{QCDSR}-\cite{MNDL} that
$\lambda_N \simeq (1 - 3)\times 10^{-2} \mbox{GeV} ^3$, and
$\sqrt{s_0} \simeq (1.1 - 1.5)\, \mbox{GeV}$. The higher values of
$\lambda_N$ and $s_0$ come from Laplace sum rules \cite{reinders},
and the lower values are from a FESR analysis \cite{MNDL} which
yields the relation $s_0^3 = 192 \pi^4 \lambda_N^2$. After fitting
Eq.(14) to the experimental data, as corrected in \cite{Brash}, we
find $\lambda_N = 0.011 \mbox{GeV}^3$, and $s_0 = 1.2
\mbox{GeV}^2$, in line with the values discussed above.
Numerically, $s_0$ is well below the Roper resonance peak, thus
justifying the model used for the hadronic spectral function. The
predicted form factor $F_1(q^2)$ is shown in Fig.2 (solid line)
together with the data. The agreement is quite impressive.
\begin{figure}
\includegraphics[scale=0.5]{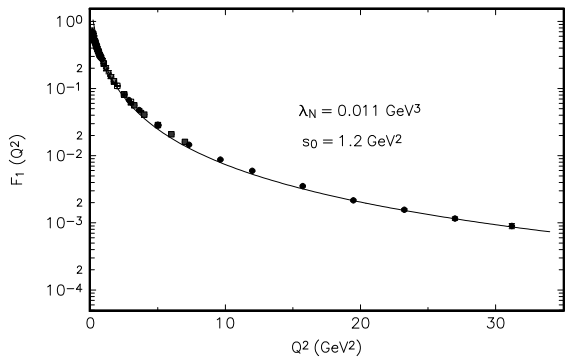}
\caption{ Theoretical results (solid line) versus corrected
experimental \cite{Brash} data on $F_{1}(Q^{2})$} \label{fig2}
\end{figure}

From the two leading power corrections in the OPE, without gluon
exchange, the one proportional to the quark condensate does not
contribute to $F_{1}(Q^{2})$, while the other, proportional to the
four-quark condensate, has a vanishing double discontinuity in the
$(s,s')$ complex plane. For details see the original article
\cite{Dom05}.

In order to extract $F_{2}(Q^{2})$, we have to consider the
leading-order non-perturbative corrections to the OPE, which in
this case corresponds to the quark condensate.

\begin{figure}
\includegraphics[scale=0.5]{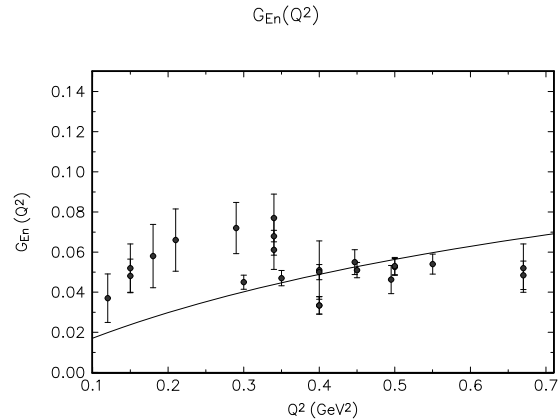}
\caption{Experimental data on $G_{E}(Q^{2})$ for the neutron
\cite{jeff}, together with the theoretical results}  \label{fig3}
\end{figure}

In the case of the proton, the contribution involving the up-quark
condensate vanishes (due to vanishing traces) and therefore we
only have a piece proportional to $\langle \bar{d} d \rangle$. Our
choice of Lorentz structure in this case is $\lh{q} \gamma^\mu$,
which appears in the QCD sector as well as in the hadronic sector
multiplying $F_{2}(Q^{2})$ but not $F_{1}(Q^{2})$. We refer to the
original article \cite{Dom05} for the full expressions. Here we
will give only the final result that emerges form the FESR
\begin{eqnarray}
F_2(Q^2)= -\frac {\langle \bar d d\rangle} { 24 \kappa_p \, M_N\,
\pi^2\,
 \lambda_N^2}
[2 s_0 ( Q^2 + s_0 ) \nonumber\\ +
   \, Q^2 \, ( Q^2 + 2 s_0 ) \,
     \ln (\frac{Q^2}{Q^2 + 2 s_0})] \; .
\end{eqnarray}
The problem is that the asymptotic behavior does not agree with
the expectations. We find
\begin{eqnarray}
   \lim_{Q^2 \rightarrow \infty}
         F_2(Q^2) = -  \frac{\langle \bar d d\rangle} { 18 \kappa_p \, M_N\, \pi^2\,
 \lambda_N^2}\nonumber\\
 \times (\frac{s_0^3}{Q^2} -  \frac{s_0^4}{Q^4} +...)
         \; ,
\end{eqnarray}
\noindent and we would expect $F_{2}(Q^{2})$ to fall faster than
$F_{1}(Q^{2})$ at least by a factor of $1/Q$ \cite{JLABF2}. A
comparison of $F_{2}(Q^{2})$ from equation $(16)$ with data shows
a disagreement at the level of a factor two, which cannot be
improved adjusting the values of $\lambda $ and $s_{0}$. The main
reason behind the disagreement is the soft $q^{2}$-dependence of
$F_{2}(Q^{2})$.

We can do the same analysis for the neutron form factors. It turns
out that $F_{1n}(Q^{2})$ for the neutron is numerically very small
and consistent with zero, except near $Q^{2}=0$ due to the
divergence in the chiral limit. Since $F_{1n}(Q^{2}) \approx 0$,
the Sachs form factor is proportional to $F_{2n}(Q^{2})$. In Fig.3
we show a plot of the electric Sachs form factor for the neutron.
At low $Q^{2}$ there is reasonable agreement with the experimental
data. However, for higher momentum transfers the disagreement
turns out to be serious due to the soft $1/Q^{2}$ behavior of
$F_{2n}(Q^{2})$.

\noindent
 {\bf Acknowledgements:}   We acknowledge support fom Fondecyt
 under grants 1051067 and 7050125

\end{document}